\def\mathform#1{\relax\ifmmode{#1}\else{$#1$}\fi}
\def\nonE{\mathform{\rlap/e}}
\documentstyle[preprint,floats,prd,aps]{revtex}
\tighten
\begin{document}
\noindent
\preprint{\vbox{
\hbox{\bf\ \makebox[3in]{} \makebox[1in]{CLNS 94/1297}}
\hbox{\bf\ \makebox[3in]{} \makebox[1in]{CLEO 94-20}}
\hbox{\bf\  \makebox[3in]{} \makebox[1in]{September 1994}}}}

\title{\Large \bf
Measurement of the branching fraction for $\Upsilon (1S)
\rightarrow \tau^+ \tau^-$}
\setcounter{page}{0}
\author{
D.~Cinabro,$^{a}$ T.~Liu,$^{a}$ M.~Saulnier,$^{a}$ R.~Wilson,$^{a}$
H.~Yamamoto,$^{a}$
T.~Bergfeld,$^{b}$ B.I.~Eisenstein,$^{b}$ G.~Gollin,$^{b}$
B.~Ong,$^{b}$ M.~Palmer,$^{b}$
% ...
}
\address{
\bigskip % Delete for submission to PRL or PRD
{\rm (CLEO Collaboration)}\\  % DO NOT Delete!
\newpage % Delete for submission to PRL or PRD
$^{a}${Harvard University, Cambridge, Massachusetts 02138}\\
$^{b}${University of Illinois, Champaign-Urbana, Illinois, 61801}\\
% ...
\bigskip % Delete for submission to PRL or PRD
}        % DO NOT delete!
\pagestyle{empty}
\maketitle
\pagestyle{plain}
\setcounter{footnote}{1}

\begin{abstract}
We have studied the leptonic decay of the $\Upsilon (1S)$
resonance into tau pairs
using the CLEO II detector. A clean sample of tau pair events is
identified via events containing two charged particles
where exactly one of the
particles is an identified electron. We find
$B(\Upsilon(1S) \rightarrow \tau^+ \tau^-) =
\left(2.61~\pm~0.12~{+0.09\atop{-0.13}}\right)\%$. The result is consistent
with expectations from lepton universality.
\end{abstract}

\footnotetext[1]{Permanent address: University of Hawaii at Manoa}

\newpage

% body of the paper goes here

\par
One of the interesting aspects of heavy quarkonia is that in the lower
energy states the electromagnetic decays compete with the strong decays
due to OZI suppression. In the $b\bar{b}$ system, the first three $\Upsilon$
resonances all lie below the threshold for strong decay into pairs of
$B$ mesons, and the measured leptonic decays are of the order of a few
percent. For the $\Upsilon (1S)$, the world average of the branching fraction
into tau pairs is ($2.97 \pm 0.35)\%$\cite{ref-PDG}
based on two measurements, one from CLEO\cite{ref-OldC}
and one from ARGUS\cite{ref-OldA}
\rlap.\ \  Comparing the tauonic decay
rate to the $e^+ e^-$ and $\mu^+ \mu^-$ rates is an
interesting test of lepton universality. The $e^+e^-$ and $\mu^+ \mu^-$
branching fractions have been measured and lie about one standard deviation
lower than the tau pair branching fraction\cite{ref-PDG}\rlap.
\par
We describe here a new measurement of the tauonic branching fraction of the
$\Upsilon (1S)$\cite{ref-HLPhD}
which is significantly more precise than the two
previous determinations. The analysis method also differs significantly
from the previous two analyses. The previous CLEO measurement identified
taus in their 1-vs.-3 topology\footnote[2]
{The terminology $a$-vs.-$b$
topology refers to a tau-pair event in which the final decay products of
one tau includes $a$ charged particles and the final state of the other
tau includes $b$ charged particles.} and the ARGUS measurement was based
upon $\Upsilon (1S)$ mesons produced via $\Upsilon (2S)$ decays. Here we use
data taken on the $\Upsilon (1S)$ resonance and look for tau pair events
where one tau has decayed into $e \nu \bar\nu$ and the other has decayed
into a final state having one charged particle which is not an electron
(we use the notation \nonE\ below to denote this track, which is a muon,
pion, or kaon). This allows us to select a very pure sample of tau pair
events and avoid uncertainties associated with the hadronic background.
The presence of the electron also improves our understanding of the
trigger efficiency of the detector.
\par
The data were recorded with the CLEO II detector which operates at the
Cornell Electron Storage Ring (CESR). The CLEO II detector is described
in detail elsewhere\cite{ref-CDet}
\rlap. \ Excellent electron identification is
provided by the charged particle tracking system and the electromagnetic
calorimeter which are inside the solenoidal magnet with a 1.5 T field.
The calorimeter consists of 7800 CsI(Tl) crystals providing excellent
energy resolution and fine segmentation.
\par
The data sample used in this analysis corresponds to an integrated
luminosity of 49 pb$^{-1}$ at the energy of the $\Upsilon (1S)$
resonance. In addition, a 101 pb$^{-1}$ data sample taken in the
continuum between the $\Upsilon (3S)$ and $\Upsilon (4S)$ resonances
is used for subtraction of the non-resonant contribution.
The trigger conditions were the same for the two datasets.
\par
In measuring the branching fraction we must make a large
subtraction to remove the contribution from the non-resonant
process $e^+ e^- \rightarrow
\tau^+ \tau^-$ from the $e^+ e^- \rightarrow \Upsilon \rightarrow \tau^+
\tau^-$ signal. Because the background continuum process is measured at a
different energy from the signal process ($E_{beam}$ of 5.263 GeV vs.
4.730 GeV for the $\Upsilon (1S)$), we try to
choose cuts for which the energy dependence of the efficiency is well
simulated by our Monte Carlo.
Determination of the trigger efficiency is critical,
so we insist that each event must fire a specific set of well-understood
hardware trigger elements.
The set of elements we use combines information from the
calorimeter and the central drift chamber to identify events with an
electron candidate and at least one other charged track.
\par
There are two aspects to selecting events: electron identification
cuts and tau topology cuts. For electron identification,
we match the charged track with energy deposited in the CsI calorimeter
and insist that this energy be consistent with the measured momentum of
the track. In particular, if
$p$ is the momentum of the charged track and
$E$ is the associated energy in the calorimeter, we require
	$0.85 \leq E/p \leq 1.1$.
Further, we require that there be at most one nearby photon which
could overlap with the electron shower and that
there be no associated signal in the muon chambers.
Tracks that fail any of these criteria are classified as ``\nonE\ ''.
\par
The tau event selection cuts are designed to accept two track tau pair
events while minimizing the contribution from other processes. Muon pair
and Bhabha events can be eliminated by the particle identification
criteria (one $e$, one \nonE\ ) and by insisting that
there be missing energy in the event. Some additional cuts are
implemented to reduce the contamination from radiative Bhabhas
(such as (d) and (j) below), hadronic events (cut (h) below),
and two-photon processes (cut (g) below).
The polar angles of the
calorimeter shower and the missing momentum in the
event are denoted $\theta_{E},$ and $\theta_{miss}$,
respectively. $E_{vis}$ is the total visible energy, charged and neutral,
in the
event. $\Sigma E$ is the summed energy of all the calorimeter showers.
The event topology cuts applied are:
\begin{description}
\item [{\rm (a)}]	Exactly two good charged tracks
with a net charge of zero.
\item [{\rm (b)}] Both tracks have good, unique matches to calorimeter showers.
\item [{\rm (c)}] One track must be called ``$e$'' and the other ``\nonE '' by
the
electron identification criteria.
\item [{\rm (d)}] At most one track with $p \geq 0.85 E_{beam}$.
\item [{\rm (e)}] Both tracks satisfy $p \geq 0.65$ GeV.
\item [{\rm (f)}] $\cos(\theta_{E}) \leq 0.707$ for the showers matched
to charged particle tracks.
\item [{\rm (g)}] $|\cos(\theta_{miss})| < 0.98$.
\item [{\rm (h)}] At most 10 isolated showers in the calorimeter.
\item [{\rm (i)}] $20\% < E_{vis} / E_{cm} < 90\%$.
\item [{\rm (j)}] $(\Sigma E / E_{cm}) < -0.36(|\vec{p}(\nonE )| /
	 E_{beam}) + 0.78$.
\item [{\rm (k)}] $|\vec{p}(e)| < 90\%\ E_{beam}$;
	$|\vec{p}(\nonE )| < 96\%\ E_{beam}$.
\end{description}

\par
These cuts select 4899 events in the $\Upsilon (1S)$ data sample, and
5824 events in the continuum sample. Figure 1a. shows the uncorrected
momentum distribution of the electron candidate in our continuum events.
Figure 1b. shows the uncorrected momentum distribution for the \nonE\
track for the same data. Figures 2a. and 2b. show the same quantities
for the $\Upsilon (1S)$ data sample. The normalized tau Monte Carlo
curves shown in the figures include the effect of the cuts, but not any
background, misidentification, or trigger efficiency effects. The good
agreement with the data, especially at higher momentum, demonstrates
that the raw samples have small background and fake contaminations.
The trigger efficiency effects are addressed below.
\par
A number of background sources were studied. Cosmic rays, $\mu$
pairs, beam-wall and beam-gas interactions were investigated using muon
chamber, timing, and vertex information. All of these backgrounds were
determined to be negligible. Similarly, no radiative Bhabha events
passed our cuts when Monte Carlo events were examined. The
most significant backgrounds could potentially
come from hadronic and two-photon
processes. No hadronic continuum Monte Carlo events ($e^+ e^-
\rightarrow q \bar{q}$) passed our cuts when a fully simulated sample
corresponding to an integrated luminosity of 230 pb$^{-1}$ was examined.
In addition, over one million $\Upsilon (1S) \rightarrow ggg$ Monte
Carlo events were studied and only one event passed our cuts. Thus we
find our hadronic background to be less than 0.02\% in our tau samples.
Two-photon Monte
Carlo data corresponding to 1299 pb$^{-1}$ with the final state $e^+ e^-
\tau^+ \tau^-$ were examined. Assuming other two-photon processes
produce a negligible background, we found 28.3 $\pm$ 1.4 two-photon
events in our continuum data, and 13.8 $\pm$ 0.7 events in the $\Upsilon
(1S)$ data. After the continuum subtraction, this background is
therefore also negligible, as the 28.3 events scale to 16.1 in the
subtraction.
\par
There is also a ``background'' from real tau events.
We take as the fraction of tau events in our signal process
$B(\tau^+\tau^- \rightarrow \ e\nonE) =
2B_e(B_1 - B_e)$, where $B_1$ is the one prong topological branching
fraction and $B_e$ is the electronic branching fraction. (We use values for
these branching fractions taken from the Particle Data Group.\cite{ref-PDG} )
Thus any 1-vs.-3 (or 1-vs.-5) tau-pair event,
or a 1-vs.-1 tau event where the $e$ or \nonE\ has been misidentified is
treated as background.
The background is most easily expressed as
a fraction of the tau-pair events which are ``fakes.'' Using
about 100,000
generic Monte Carlo tau-pair events,
we found a fake rate of
$(4.1 \pm  1.5)\%$  for the $\Upsilon (1S) \rightarrow \tau^+ \tau^-$
events. The fake rates for
continuum tau production at the beam energies of our two data samples
differ only slightly from this and from each other.
\par
In calculating this fake rate we used data to determine the fraction of
hadrons passing the critical cut, $0.85 < E / p < 1.1$. Muons are
very unlikely to deposit this fraction of their energy in the
electromagnetic calorimeter, but hadrons occasionally do. Tracks from
$K^0_s$ decays in hadronic events were used to measure, as a function of
momentum, the fraction of pions which satisfy our $E / p$ cut. The
$\pi^+$ and $\pi^-$ have differing fractions and were treated separately
in our analysis. (We assume that any differences between kaons and pions
are negligible for this purpose since there are approximately twenty times
more pions than kaons in our tau events.) A very clean sample of $K^0_s
\rightarrow \pi^+ \pi^-$ candidates was selected by taking oppositely
charged tracks from a secondary vertex and requiring that the $K^0_s$
momentum vector point back to the primary vertex. Pion candidates were
chosen from those $K^0_s$ candidates with an invariant mass between 0.48
and 0.51 GeV/c$^2$. If the $p \pi$ invariant mass was consistent with the
process $\Lambda \rightarrow p \pi^-$, the tracks were rejected to avoid
proton contamination. The calculated event selection efficiency for tau-pair
events from $\Upsilon (1S)$ decay is $(26.69 \pm 0.21)\%$; this value
also depends somewhat upon these fake rates.
\par
Trigger efficiencies were determined from the data.
The CLEO II trigger has been described in detail
elsewhere\cite{ref-Trgo}\rlap.\ \
There are three levels to the hardware trigger and
each level has several elements
associated with different subsystems of the CLEO II detector. We
required in our event selection a
specific set of four elements, each coming from a different detector
subsystem: time-of-flight, barrel calorimeter, central drift chamber,
and vertex detector.
All trigger element efficiencies were studied using radiative Bhabha
events except for the time-of-flight efficiency which was studied in the tau
events themselves. Radiative Bhabha events were selected from the
same $\Upsilon (1S)$ dataset.
The strategy was to look at events which fired a different trigger
line which does not involve the detector subsystem under study and measure the
fraction of those events that fire the given element.
The efficiencies of all but the calorimeter element are found to be
independent of momentum in the momentum
region of interest (that is, the region $p > 0.60$ GeV/c).
The efficiency of the calorimeter element is
100\% when the momentum of the electron is greater than 0.95 GeV/c,
but decreases with decreasing momentum below that value, down to an
efficiency of 86\% at $p = 0.65$ GeV/c, the lowest momentum allowed by our
selection criteria. We obtained a function giving the trigger efficiency
versus momentum by fitting our data. Then
taking the fitted function and a Monte Carlo generated momentum
distribution for the electrons in our tau events, we calculated a
momentum integrated trigger efficiency. Momentum
distributions were generated for continuum produced taus at beam
energies of 5.263 GeV and 4.730 GeV and for
$\Upsilon (1S) \rightarrow \tau^+ \tau^-$.
The three momentum distributions are slightly different and give
slightly different efficiencies.
For $\Upsilon (1S) \rightarrow \tau^+ \tau^-$, the trigger efficiency is
$(95.1 \pm 1.2)\%$.
The error was
determined from the statistical errors on the measurements and
the spread in values obtained by using different
fitting functions.
The ratio of the continuum trigger efficiencies at the two beam
energies ($\epsilon^{4.730}/\epsilon^{5.263}$)
was seen to be quite insensitive to the
method of integration over momentum, and is found to be 0.998 $\pm$
0.002.
\par
To calculate the number of tau pair events from $\Upsilon (1S)$ decays,
$N_{\tau}$, we subtract fakes from the raw tau event yields, scale the
continuum value (by relative luminosities, cross sections, and
efficiencies), subtract it from the $\Upsilon (1S)$ value, and divide by
the efficiencies and $B(\tau^+\tau^- \rightarrow \ e\nonE).$
We obtain:
$N_{\tau} = 25100 \pm 1300 \pm 800.$ We calculate that 31\% of the tau
pairs observed at 4.730 GeV are from the $\Upsilon (1S)$ resonance.
The quoted systematic error comes from the uncertainties in the trigger
efficiencies and fake rates discussed previously, the uncertainties in
$B_1$ and $B_e$, a 0.5\% error in the ratio of
luminosities\cite{ref-LUM}, and uncertainties in the ratio of continuum
tau-pair cross sections at the two beam energies. The cross sections
were calculated from $\alpha^3$ QED\cite{ref-KORALB}.
We found the ratio of the cross
sections to be independent of the cutoff energy of the photon and to
have an uncertainty of 0.5\%\cite{ref-AJWqed}.
We also studied the sensitivity of the
result to the lower bound on the track momenta by varying that
cut from 0.65 GeV to 1.0 GeV. This caused $N_{\tau}$ to change by less
than 0.5\%.
\par
We normalize the number of tau events to the number of hadronic events
from the $\Upsilon (1S)$. Hadronic events were selected by requiring
three or more charged tracks, $E_{vis} \geq 0.30 E_{cm}$,
and $\Sigma E \geq 0.12 E_{cm}$. The primary event vertex was also
required to be consistent with the beam position. From Monte Carlo,
we find the efficiency for hadrons from the $\Upsilon (1S)$ resonance is
98.8\% with these cuts. The efficiency for continuum hadronic events is
94.0\% at the $\Upsilon (1S)$ energy and 94.4\% at $E_{beam} =$ 5.263
GeV. The continuum subtracted hadronic event yield is found to be
$N_{hadron} = 886600 \pm 1100$ events where the error is statistical
only. A Monte Carlo calculated tau background of 7300 events was
subtracted to arrive at this number. The dominant systematic error comes
from the determination of the efficiency for hadrons from the $\Upsilon
(1S)$. We examined the agreement between the Monte Carlo simulation and
the continuum subtracted hadronic data. The data show an excess at low
multiplicity, leading to a possible overestimation of the efficiency.
Because the discrepancy tends to produce a
lower efficiency and because one cannot have an efficiency in excess of
100\%, we assign an asymmetric systematic error of
${+11,200\atop{-36,000}}$ events to the number of hadrons.
\par
In obtaining our branching fraction, we first define ${\bar{B}}$ as the
ratio of $N_{\tau}$ to $N_{hadron}$. Then, assuming lepton universality:
\begin{equation}
B(\Upsilon(1S) \rightarrow \tau^+ \tau^-) = { \bar{B} \over {1 +
3\bar{B} }}. \end{equation}
Using the numbers of taus and hadrons we observed gives: $
{\bar{B}} = 0.0283 \pm 0.0014 {+0.0010 \atop {-0.0015}}$, which yields:
\begin{equation}
$$B(\Upsilon(1S) \rightarrow \tau^+ \tau^-) =
\left(2.61~\pm~0.12~{+0.09\atop{-0.13}}\right)\%.$$\end{equation}
We can test
lepton universality
by comparing this number to previous measurements of the mu-pair and
electron-pair branching fractions. These values are\cite{ref-PDG}\rlap: \
$B(\Upsilon (1S) \rightarrow \mu^+ \mu^-) = (2.48 \pm 0.07)\%$, and
$B(\Upsilon (1S) \rightarrow e^+ e^-) = (2.52 \pm 0.17)\%$.\footnote[3]
{One can also obtain a branching fraction by assuming only electron-muon
universality. Using the muon pair branching fraction quoted above we
obtain $B(\Upsilon(1S) \rightarrow \tau^+ \tau^-) =
\left(2.62~\pm~0.13~{+0.09\atop{-0.13}}\right)\%.$}
We see that
our measurement is in closer agreement with lepton universality than the
previous world average.
\par
In conclusion, we have measured the tauonic branching fraction of the
$\Upsilon (1S)$ resonance. The value we obtain is more precise than
previous measurements and is consistent with expectations from lepton
universality.
\par
%\centerline{\bf ACKNOWLEDGEMENTS}
%\smallskip
We gratefully acknowledge the effort of the CESR staff in providing us with
excellent luminosity and running conditions.
J.P.A. and P.S.D. thank
the PYI program of the NSF, I.P.J.S. thanks the YI program of the NSF,
G.E. thanks the Heisenberg Foundation,
I.P.J.S., and T.S. thank the TNRLC,
K.K.G., M.S., H.N.N., J.D.R., T.S.  and H.Y. thank the
OJI program of DOE
and P.R. thanks the A.P. Sloan Foundation for
support.
This work was supported by the National Science Foundation, the
U.S. Dept. of Energy and the Natural Sciences and Engineering Research
Council of Canada.
%\newpage

\newpage
\begin{figure}
%\centerline{\psfig{figure=fig1.eps,height=6.5in}}
\caption{Momentum distribution of a) the electron candidate and
b) the \nonE\ candidate in the
continuum data sample. The data are
shown by the solid circles and the histogram is from a tau
Monte Carlo simulation.}
\end{figure}
\newpage
\begin{figure}
%\centerline{\psfig{figure=fig2.eps,height=6.5in}}
\caption{Momentum distribution of a) the electron candidate and
b) the \nonE\ candidate in the
$\Upsilon(1S)$ data sample. The data are
shown by the solid circles and the histogram is from a tau
Monte Carlo simulation.}
\end{figure}

\begin{thebibliography}{9}
\bibitem{ref-PDG}
Particle Data Group, L. Montanet {\it et al.}, Phys. Rev. {\bf D50}
(1994) 1173.
\bibitem{ref-OldC}
CLEO Collab., R. Giles {\it et al.}, Phys. Rev. Lett. {\bf 50} (1983) 877.
\bibitem{ref-OldA}
ARGUS Collab., H. Albrecht {\it et al.}, Phys. Lett. {\bf 154B} (1985) 452.
\bibitem{ref-HLPhD}
Ha Lam, Ph. D. thesis, University of Kansas (1994), unpublished.
\bibitem{ref-CDet}
CLEO Collab., Y. Kubota {\it et al.}, Nucl. Instr. and Meth. {\bf A320}
(1992) 66.
\bibitem{ref-Trgo}
C. Bebek {\it et al.}, Nucl. Instr. Meth. {\bf A302} (1991) 261.
\bibitem{ref-LUM}
B. Heltsley, private communication;
G. Crawford, {\it et al.}, Nucl. Instr. and Meth. {\bf A345 } (1994) 429.
\bibitem{ref-KORALB}
{\tt KORALB 2.1/ TAUOLA 1.5}:
S.~Jadach and Z.~Was,
{\em Comput.~Phys.~Commun.}~{\bf 36}, 191 (1985) and
{\em ibid}, {\bf 64}, 267 (1991);
S.~Jadach J.H.~K\"{u}hn, and Z.~Was,
{\em Comput.~Phys.~Commun.}~{\bf 64}, 275 (1991),
{\em ibid}, {\bf 70}, 69 (1992),
{\em ibid}, {\bf 76}, 361 (1993).
\bibitem{ref-AJWqed}
Alan Weinstein, private communication.

\end{thebibliography}
\end{document}